\documentclass{ws-procs975x65}

\begin{document}

\title{POLARIZATION PROFILES FOR SELECTED MICROLENSING EVENTS TOWARDS THE GALACTIC BULGE}

\author{A.A. NUCITA$^*$, G. INGROSSO, F. De PAOLIS, F. STRAFELLA}
\address{Department of Mathematics and Physics {\it Ennio De Giorgi}, University of Salento, and INFN via per Arnesano73100Lecce Italy \\
$^*$E-mail: nucita@le.infn.it}

\author{S. CALCHI-NOVATI}
\address{Dipartimento di Fisica ``E. R. Caianiello'', Universit\`a di 
Salerno, Via Ponte don Melillo, 84084 Fisciano (SA), Italy\\
Istituto Internazionale per gli Alti Studi Scientifici (IIASS), 
Vietri Sul Mare (SA), Italy}

\author{Ph. JETZER}
\address{Institute for Theoretical Physics, University of  Z\"{u}rich, Winterthurerstrasse 190, CH-8057 Z\"{u}rich, Switzerland\\}

\author{A. F. ZAKHAROV}
\address{Institute of Theoretical and Experimental Physics,B. Cheremushkinskaya 25, 117259 Moscow,  Russia \\
Bogoliubov Laboratory of Theoretical Physics, Joint Institute for Nuclear Research, 141980 Dubna, Russia}

\begin{abstract}
Gravitational microlensing, in case of relevant finite source size effects,  provides an unique tool for the study of stellar atmospheres 
through the enhancement of a characteristic polarization signal. Here, we consider a set of highly magnified events
and show that for 
different types of source stars (as hot, late type main sequence
and cool giants) showing that the polarization strength may be of $\simeq 0.04$ 
percent for late type stars and up to a few percent
for cool giants.

\end{abstract}

\bodymatter

\section{Introduction}\label{aba:sec1}

Gravitational microlensing, initially developed to search 
for MACHOs in {the} Galactic halo and near the Galactic disc 
\cite{Pacz86}\cite{Macho93}\cite{Eros93}\cite{Ogle93}, has become nowadays a powerful method to
test the stellar atmosphere models and to study of the star limb-darkening
profile, i.e. the variation of the intensity 
from the disc center to the limb. Furthermore, this technique allowed the discovery of
exoplanetary systems by observing deviations in the 
light-curves expected for single-lens events \cite{Dominik10}\cite{Gaudi10}. 
Microlenses can also spatially resolve a source star thanks to caustic structures 
created by a lens system \cite{SEF}. Caustics are formed by a set of closed curves, along which the 
point source magnification is formally infinite, with a steep increase in 
magnification in their vicinity. 

The aim of the present work is to consider the polarization variability of the source star light 
for real events, taking into account 
different polarization  mechanisms according to the source star type.
Indeed, variations in the polarization curves are similar to finite source 
effects in microlensing when color effects may appear due to 
limb darkening and color distribution across the disc. However, 
the light received from the stars is usually unpolarized, 
{since the flux from each stellar disc element is the same.}
A net polarization of the stellar light may be introduced by some
suitable asymmetry in the stellar disc as eclipses, tidal distortions, 
stellar spots, fast rotation and magnetic fields or also in the propagation 
through the interstellar medium.

In the case of microlensing events, polarization in the stellar light may be induced
by the proper motion of the lens star through the source star
disc. In such case, different parts are magnified
by different amounts giving rise not only to a time dependent gravitational magnification 
of the source star light but also to a time dependent polarization.

Good candidates for polarization observationsw ould be events microlensing 
involving young, hot giant star sources. Indeed, these objects
have electron scattering atmospheres needed for producing limb 
polarization through Thomson scattering \cite{Simmons95b}.
However, the bulge of the Galaxy does not contain a large number of
hot giant stars. However, polarization may be also induced by the scattering of star light off 
atoms, molecules and dust grains in the adsorptive atmospheres of evolved, 
cool stars as shown by Refs. \refcite{Simmons02} and \refcite{Ignace06}.
These objects, that do not have levels of polarization as high 
as those predicted by the Chandrasekar model, may display an intrinsic 
polarization of up to several { percent}, 
due to the presence of stellar winds that give rise to extended adsorptive envelopes. 
This is the case for many cool giant stars, in particular for the red giants. 
Such evolved stars constitute a significant fraction of the lensed sources 
towards the Galactic bulge, the LMC \cite{Alcock97} and the M31 galaxy\cite{Sebastiano10} 
making them valuable candidates for observing variable 
polarization during microlensing events.

In Ref. ~\refcite{ingrosso2012} (but see also the references therein for more details)
we calculated the polarization profile as a function of the time for a 
selected sample of both single events (11 highly magnified single-lens cases with identified source star type) 
and 6 exoplanetary microlensing events observed towards the Galactic bulge. In predicting the polarization light curves of each event we considered 
the nature of the source star, i.e. a late type main sequence and/or cool giant stars. Indeed, different polarization mechanisms 
take place in the stellar 
atmospheres, depending on the source star type: 
photon (Thomson) scattering on free electrons, coherent (Rayleigh) scattering
off atoms and molecules, and photon scattering on dust grains, 
{ for hot, late type and cool giant stars} (with
extended atmospheres), respectively.

We note that high magnification (up to 12 mag in I band) events, the expected polarization signal 
can reach values as high as 0.04 percent at the peak in the case of 
{late type source stars} and up to a few {percent} in the case of cool giant source stars 
(red giants) with extended envelopes. 
For these events the primary lens crosses the  source star disc 
({\it transit} events) and relatively large values of $P$ are thereby produced
due to large finite source effects and the large magnification
gradient throughout the source star disc, with the time duration of the polarization peak 
varying from 1 h to 1 day (depending on the source star radius and the lens impact parameter).
Similar values of polarization may also be obtained
in exoplanetary events when the source star crosses the primary or the
planetary caustics. While in the former case (as for single-lens 
events) the peak of the polarization signal always occurs at symmetrical 
points with respect to the instant $t_0$ of maximum magnification, in the latter
case the polarization signal may occur at any (and generally unpredictable)
time during the  event. 
 
As a last remark, we note that the available
instrumentation may already detect a polarization signal down to a degree of few percent: 
the FORS2 camera on the ESO VLT telescope is capable to measure 
the polarization signal for a 12 mag
source star with a precision of 0.1 {percent} in 10 min integration time, 
and for a 14 mag star in a 1h. Hence, polarization measurements in
highly magnified microlensing 
events may offer the unique opportunity
to probe stellar atmospheres of Galactic bulge stars and, given sufficient observational precision, 
may in principle provide independent constraints on the lensing parameters also for exoplanetary events.

\end{document}